\documentclass[journal=ancac3,manuscript=article]{achemso}

\usepackage[version=3]{mhchem} 
\usepackage[hidelinks]{hyperref}
\usepackage{xcolor}

\newcommand*{\rttensortwo}[1]{\bar{\bar{#1}}}

\author{Jessica Meier}
\altaffiliation{These authors contributed equally}
\author{Luka Zurak}
\altaffiliation{These authors contributed equally}
\affiliation[First University]
{Nano-Optics and Biophotonics Group, Experimental Physics 5, Institute of Physics, University of Würzburg, Germany}
\author{Andrea Locatelli}
\affiliation[Second University]
{Department of Information Engineering, University of Brescia, Italy}
\author{Thorsten Feichtner}
\author{René Kullock}
\author{Bert Hecht}
\email{hecht@physik.uni-wuerzburg.de}
\affiliation[First University]
{Nano-Optics and Biophotonics Group, Experimental Physics 5, Institute of Physics, University of Würzburg, Germany}

\title{Controlling field asymmetry in nanoscale gaps for second harmonic generation}


\begin{document}


\begin{abstract}
  Plasmonic dimer antennas create strong field enhancement by squeezing light into a nanoscale gap. These optical hotspots are highly attractive for boosting nonlinear processes, such as harmonic generation, photoelectron emission, and ultrafast electron transport. Alongside large field enhancement, such phenomena often require control over the field asymmetry in the hotspot, which is challenging considering the nanometer length scales. Here, by means of strongly enhanced second harmonic generation, we demonstrate unprecedented control over the field distribution in a hotspot by systematically introducing geometrical asymmetry to the antenna gap. We use focused helium ion beam milling of mono-crystalline gold to realize asymmetric-gap dimer antennas in which an ultra-sharp tip with 3$\,$nm apex radius faces a flat counterpart, conserving the bonding antenna mode and the concomitant field enhancement at the fundamental frequency. By decreasing the tip opening angle, we are able to systematically increase both field enhancement and asymmetry, thus enhancing second harmonic radiation to the far-field, which is nearly completely suppressed for equivalent symmetric dimer antennas. Combining these findings with second harmonic radiation patterns as well as quantitative nonlinear simulations, we further obtain remarkably detailed insights into the mechanism of second harmonic generation at the nanoscale. Our results open new opportunities for the realization of novel nonlinear nanoscale systems, where the control over local field asymmetry in combination with large field enhancement is essential to create nonreciprocal functionalities.
\end{abstract}

\section{Introduction}
Plasmonic nanostructures allow to squeeze light into nanoscale volumes providing strong field hotspots that can significantly enhance nonlinear effects~\cite{kauranen_nonlinear_2012, dombi_strong-field_2020}. A variety of such processes not only require high field enhancement, but also control over the symmetry of the field distribution. This includes even-order harmonic generation~\cite{bonacina_harmonic_2020}, directed electron transport~\cite{rybka_sub-cycle_2016,ludwig_sub-femtosecond_2019}, optical rectification~\cite{ward_optical_2010,piltan_optical_2017,dasgupta_optical_2018}, and photoelectron emission~\cite{karnetzky_towards_2018,zimmermann_toward_2019,shi_femtosecond_2021}. For the latter, asymmetry was geometrically imposed, however, using rather large feature sizes ($>20\,$nm), which provide field asymmetry only at the expense of drastically reduced field enhancement. On the sub-$10\,$nm scale, controlling field asymmetry is extremely challenging, and is typically randomly generated as in electromigrated gaps~\cite{heersche_situ_2007,stolz_nonlinear_2014}. Another means to break the field symmetry in a nanoscale gap are asymmetric few-cycle phase-stable laser pulses~\cite{rybka_sub-cycle_2016,ludwig_sub-femtosecond_2019}, which offer some flexibility but are experimentally challenging.

Second harmonic generation (SHG) is a phenomenon that is especially sensitive to both field enhancement and asymmetry. For centrosymmetric materials such as gold, it only occurs at metal-dielectric interfaces where the symmetry is broken~\cite{butet_optical_2015,boyd_nonlinear_2019}. However, even though symmetric plasmonic dimer antennas with nanometer-sized gaps feature among the largest reported field enhancements for bonding mode resonances~\cite{biagioni_nanoantennas_2012, hasan_nonlinear_2014}, typically axial or inversion symmetry in the overall geometry of nanostructures leads to destructive interference of the resulting SH radiation in the far-field~\cite{finazzi_selection_2007,berthelot_silencing_2012}. Recent work has addressed this so-called silencing effect by breaking the \emph{global} symmetry of either the structure geometry~\cite{canfield_local_2007,zhang_three-dimensional_2011,black_tailoring_2015,czaplicki_second-harmonic_2015,liu_enhancement_2022} or by employing light-induced symmetry breaking~\cite{li_light-induced_2021}. In addition, in some experiments a resonance at the SH was achieved~\cite{thyagarajan_enhanced_2012,aouani_multiresonant_2012,celebrano_mode_2015,gennaro_interplay_2016, liu_polarization-independent_2016}. However, in all these cases the lack of global symmetry considerably reduces the field enhancement. 

Here, we demonstrate controlled \emph{local} symmetry breaking of the field distribution in a sub-10$\,$nm gap of a plasmonic dimer antenna, while preserving the bonding mode at the fundamental frequency, which is responsible for the large field enhancement. Unprecedented control over the gap geometry is achieved using helium-based focused ion beam milling (He-FIB) of mono-crystalline gold microplatelets. We realize asymmetric-gap antennas featuring ultra-sharp tips with a tip apex radius down to $3\,$nm for a gap size of $8\,$nm, where the degree of asymmetry can be directly controlled by modifying the tip opening angle. Local symmetry breaking leads to an asymmetric SH surface polarization in the gap, which efficiently radiates into the far-field with a nonlinear coefficient of up to $\gamma_\text{SH}=1.7\times10^{-10}\,\text{W}^{-1}$. Importantly, our approach retains the full geometric freedom afforded by in-plane antenna structures. We are thus able to record SH radiation as a function of the degree of local asymmetry, and, above that, to compare a large number of asymmetric-gap antennas to their symmetric counterparts with equal bonding mode resonances. By further recording SH radiation patterns we infer a quantitative description of the microscopic origin of SHG, which confirms the experimentally achieved large degree of asymmetry of the field distribution in the antenna hotspot.

\section{Results and discussion}
\subsection{Concept of local symmetry breaking}
 
\noindent The emission of SH radiation by a plasmonic nanoantenna to the far-field is described by the total SH polarization

    \begin{align}
    \textbf{P}^{(2\omega)} = \textbf{P}_\text{S}^{(2\omega)}+\textbf{P}_\text{R}^{(2\omega)},
    \label{eq:P_SH}
    \end{align}
    
\noindent where $\textbf{P}_\text{R}^{(2\omega)}$ denotes the response polarization of the antenna at the SH frequency driven by the SH source $\textbf{P}_\text{S}^{(2\omega)}$. For centrosymmetric materials, such as gold, $\textbf{P}_\text{S}^{(2\omega)}$ is a pure surface polarization, which is  approximated as (see also Supplementary Section~1.3):

	\begin{align}
	\textbf{P}_\text{S}^{(2\omega)}(\textbf{r})=\epsilon_0 \chi^{(2)}_{\text{S},\perp,\perp,\perp}E_\perp^{(\omega)}(\textbf{r}) E_\perp^{(\omega)}(\textbf{r}) \hat{r}_\perp\delta(\textbf{r}-\textbf{r}_\text{S}),
	\label{eq:P_s}
	\end{align}
	
where $\chi^{(2)}_{\text{S},\perp,\perp,\perp}$ is the dominant component of the $\rttensortwo{\chi}^{(2)}_\text{S}$ tensor,  $E_\perp^{(\omega)}$ is the normal component of electric field at the metal surface at frequency $\omega$, $\textbf{r}_\text{S}$ denotes a position vector pointing at the surface, and $\hat{r}_\perp$ is the surface unit normal vector~\cite{reddy_revisiting_2017,wang_surface_2009,bachelier_origin_2010}.

\begin{figure*}
	    \centering
	    \includegraphics[width=.99\linewidth]{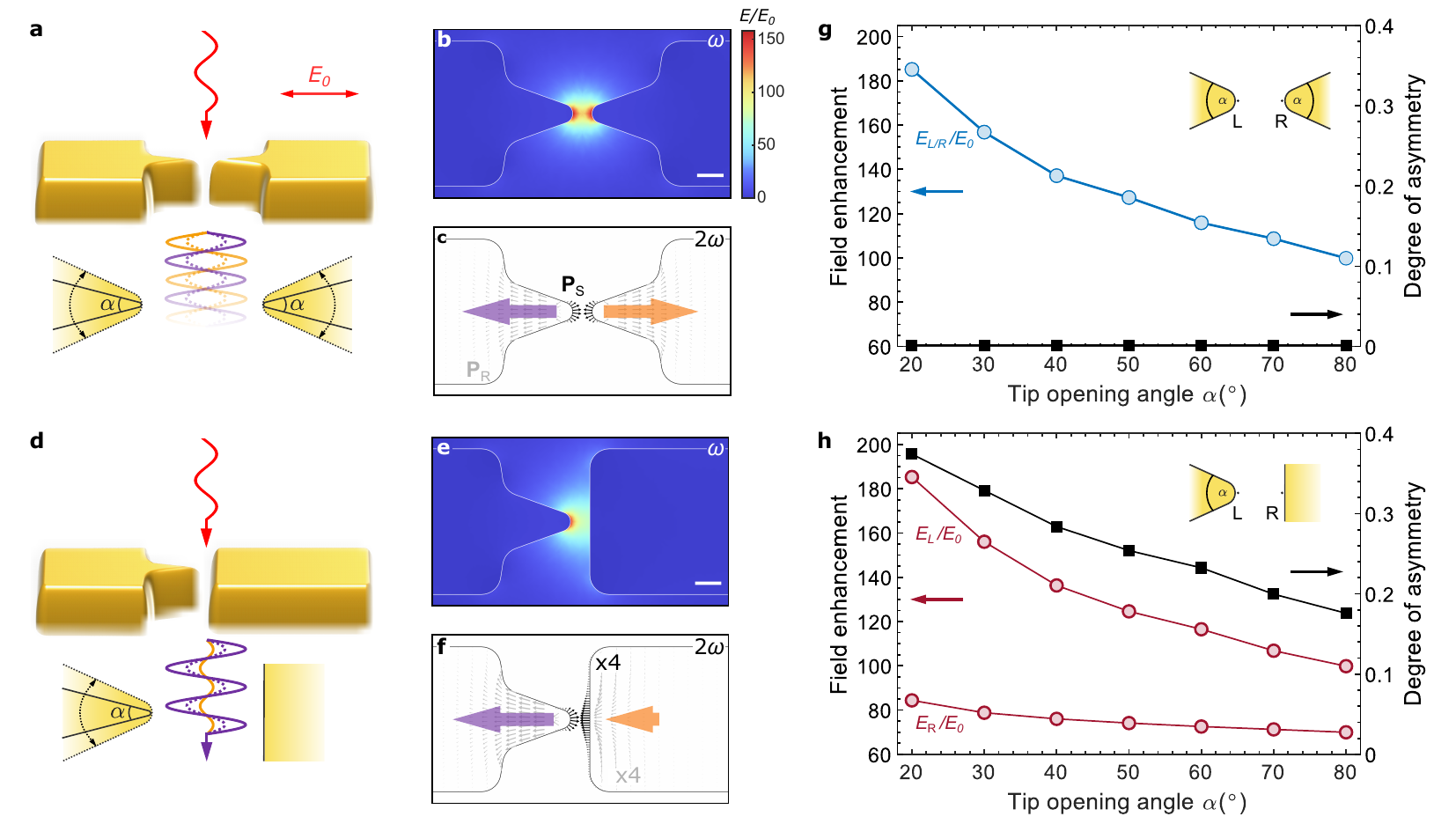}
	    \caption{\textbf{Effect of local symmetry breaking on the SHG process in gold nanoantennas.} Upon excitation with a linear polarized laser at frequency $\omega$, the SH efficiency depends on the gap geometry. \textbf{a,}~Symmetric-gap antenna. The SH response at $2\omega$ can be described by two out-of-phase oscillating effective dipoles of equal amplitude originating at the tips of the left (purple) and right (orange) antenna rod. \textbf{b,}~Simulated linear field enhancement plot evaluated around the gap region (cut parallel to the substrate at half of the antenna height) at the resonance frequency of the global bonding antenna mode. \textbf{c,}~Real part of simulated second-order polarization $\textbf{P}^{(2)}$, consisting of source polarization $\textbf{P}_\text{S}^{(2)}$ (black arrows) and response polarization $\textbf{P}_\text{R}^{(2)}$ (grey arrows), see Eq.~\eqref{eq:P_SH}. Purple and orange arrows correspond to effective dipoles introduced in \textbf{a}. \textbf{d-f,}~Asymmetric-gap antenna. The corresponding effective dipoles differ in amplitude and oscillate in-phase. The SH signal can be tuned by modifying the tip opening angle (see dashed line in the lower panel of \textbf{d}). \textbf{g,h,} Field enhancement evaluated $0.1\,$nm from the surface at positions L/R (see inset) for symmetric-gap (blue) and asymmetric-gap (red) antenna, respectively. Black square symbols correspond to the degree of asymmetry, see Eq.~\eqref{eq:DOA}. Scale bars in \textbf{b} and \textbf{e}, $10\,$nm.}
	    \label{fig:fig1}
	    \end{figure*}

	 For a mirror-symmetric plasmonic dimer antenna with nanometer-sized gap, dipolar SH radiation is expected to cancel due to the silencing effect \cite{finazzi_selection_2007,berthelot_silencing_2012}, despite the strong field enhancement in the gap at the fundamental frequency (FF) (see Fig.~\ref{fig:fig1}a and b). The silencing effect for such a symmetric-gap antenna becomes apparent when analyzing the SH polarization as displayed in Fig.~\ref{fig:fig1}c, showing both contributions, \textit{i.e.}~$\textbf{P}_\text{S}^{(2\omega)}$ (black arrows) and $\textbf{P}_\text{R}^{(2\omega)}$ (grey arrows), obtained from nonlinear simulations that will be introduced later. Integrating the nonlinear polarization over each antenna arm yields effective SH dipole moments, represented by the large purple and orange arrows for the left and right antenna arm, respectively. As these effective SH dipoles oscillate out-of-phase, SH radiation interferes destructively in the far-field.
	 
	 The silencing effect can be overcome by breaking the local symmetry of the field distribution. To this end, we taper only one side of the gap to produce a sharp tip facing a flat counterpart (see Fig.~\ref{fig:fig1}d), which yields an asymmetric field enhancement at the two sides of the gap. Since symmetry is broken locally, the global symmetry of the resonant bonding antenna mode remains unaffected resulting in a field enhancement at the FF comparable to that of the symmetric-gap antenna (Fig.~\ref{fig:fig1}e). The asymmetric field distribution at the FF leads to an asymmetric nonlinear polarization distribution visualized in Fig.~\ref{fig:fig1}f, where effective dipoles no longer cancel but oscillate in-phase, leading to strong SH radiation in the far-field.
  
  Field enhancement and field distribution in the gap strongly depend on the gap size as well as on the size of the tip apex. By keeping the gap size fixed and changing only the tip opening angle $\alpha$, \textit{i.e.}~the sharpness of the tip, we systematically modify the properties of the field. Fig.~\ref{fig:fig1}g and h display the field enhancement at the FF evaluated $0.1\,$nm from the left and right surface of the gap at half of the antenna height (labeled as positions L/R in the insets) for tip opening angles ranging from 80° down to 20° for symmetric-gap and asymmetric-gap antennas, respectively. For the same tip opening angles symmetric-gap and asymmetric-gap antennas exhibit an approximately equal field enhancement of about 100 for the case of 80°, which increases up to 185 for a tip opening angle of 20°. To assess the influence of field asymmetry on SHG, we define the degree of asymmetry (DOA) of the system as the contrast between field enhancements at positions L and R:
  
  \begin{align}
    \text{DOA}:=\frac{E_\text{L}-E_\text{R}}{E_\text{L}+E_\text{R}}.
    \label{eq:DOA}
  \end{align}

\noindent If we assume dipole-like sources situated at the left and right side of the gap with strengths proportional to square of the field amplitudes (see Eq.~\ref{eq:P_SH}), the expected SH power radiated to the far-field is proportional to the square of the effective dipole and can be written as

\begin{align}
P_\text{SH}\propto[E_\text{L}^2-E_\text{R}^2]^2 \propto E_\text{avg}^4\cdot \text{DOA}^2,
\label{eq:P_SH_DOA}
\end{align}

\noindent where $E_\text{avg}=(E_\text{L}+E_\text{R})/2$ is the average of the field strengths to the left and to the right side of the gap. Eq.~\eqref{eq:P_SH_DOA} highlights the importance of both field enhancement and asymmetry of the system with respect to achieving strong SHG. Although both types of antenna exhibit comparable field enhancement, the DOA is vanishing for all tip opening angles in the case of symmetric-gap antennas (black line in Fig.~\ref{fig:fig1}g) implying complete suppression of SHG. For asymmetric-gap antennas, however, modifying the tip opening angle allows us to separately change the field enhancement at the two sides of the gap and consequently to achieve non-zero DOA (Fig.~\ref{fig:fig1}h). As the tip gets sharper both field enhancement and DOA are increased, thus leading to strongly enhanced SHG. 
	
\subsection{Fabrication and optical characterization}

	\begin{figure*}
	    \centering
	    \includegraphics[width=.5\linewidth]{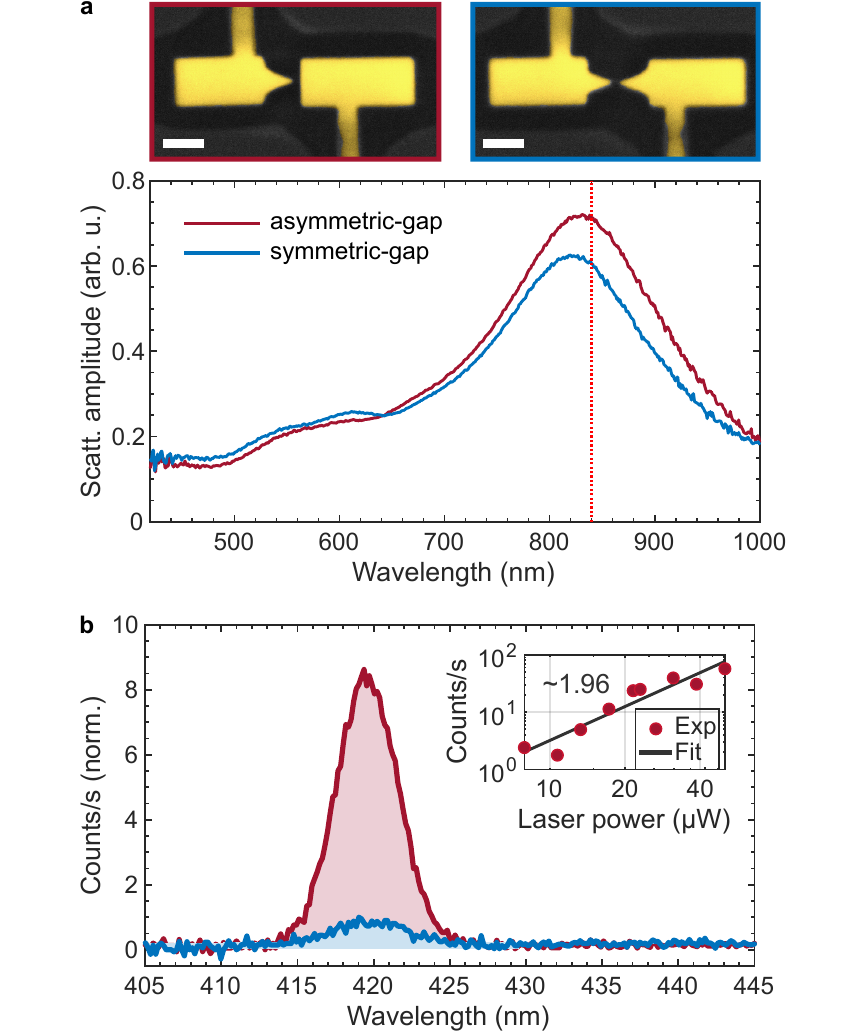}
	    \caption{\textbf{Influence of the local gap geometry on the SH radiation.} \textbf{a,} Top panel:~Colored SEM images of an asymmetric-gap and symmetric-gap antenna, respectively. Both antennas have a gap size of $9\,$ nm and tip opening angles of 40°. Scale bars, $50\,$nm. Bottom panel:~Linear white-light scattering spectra. Antennas were designed such that their resonances match the excitation wavelength of $840\,$nm. A small shoulder at around $550\,$nm appears for both antennas, which is attributed to the electrical connection wires. For the symmetric-gap antenna, a second peak at $600\,$nm corresponding to the antibonding mode appears. \textbf{b,} SH radiation spectra. Integration time was set to $60\,$s. The inset shows a bi-logarithmic plot of the power dependence of an exemplary asymmetric-gap antenna confirming the quadratic dependence of the SH signal on the pump power.}
	    \label{fig:fig2}
	    \end{figure*}
    
\noindent Realizing local symmetry breaking in plasmonic nanoantennas requires full control over the shape of the narrow gap region that can hardly be achieved even with high-end nanofabrication techniques, such as electron-beam lithography~\cite{chirumamilla_3d_2014}, electromigration~\cite{park_fabrication_1999,stolz_nonlinear_2014}, break junctions~\cite{xiang_mechanically_2013,laible_flexible_2020}, metal-nanoparticle assemblies~\cite{baumberg_extreme_2019,lim_nanogap-engineerable_2010,li_light-induced_2021}, and focused ion beam milling of mono-crystalline gold microplatelets with gallium ions (Ga-FIB)~\cite{huang_atomically_2010}.
He-FIB of evaporated gold films already surpasses pure Ga-FIB and reaches sub-$10\,$nm precision~\cite{kollmann_toward_2014}. However, the combination of He-FIB with mono-crystalline gold microplatelets, results in even smaller feature sizes leading to an unprecedented fabrication accuracy and reproducibility. We use a three-step milling approach: First Ga-FIB is used to create a rough outline of the antenna, followed by He-FIB to precisely define the antenna shape and finally the gap (see Supplementary Fig.~S7). With this combined Ga-/He-FIB approach we are able to realize asymmetric-gap antennas with ultra-sharp tips exhibiting a radius of curvature at the apex down to $3\,$nm for a gap size of $8\,$nm (see Supplementary Section 2.4).

To demonstrate the influence of local symmetry breaking on SHG, we first analyze the SH radiation for two exemplary asymmetric-gap and symmetric-gap antennas with a tip opening angle of 40°. Antennas were fabricated from the same $40\,$nm-thick gold platelet on a glass substrate and exhibit equal dimensions in the gap region, confirmed by scanning electron microscopy (SEM, top panel of Fig.~\ref{fig:fig2}a). Electrical connection wires (vertical structures in Fig.~\ref{fig:fig2}a) were included in the design to reduce charging effects during FIB milling and SEM characterization, while hardly affecting the optical response of the antenna~\cite{prangsma_electrically_2012}. Linear scattering spectra (bottom panel in Fig.~\ref{fig:fig2}a) show that both antennas are resonant at $840\,$nm, which matches the excitation wavelength of the pump laser.

For SH experiments, a pulsed $100\,$fs titanium-sapphire laser centered at $840\,$nm is focused with an oil-immersion microscope objective (NA $1.45$) through the glass substrate onto individual antennas with the polarization parallel to the long axis of the antenna. Broadening of the laser pulse due to dispersion is precompensated by a prism pulse compressor before the objective (see Supplementary Section~2.2). SH radiation is collected by the same objective and analyzed with a spectrometer after filtering out the pump light.

The recorded spectra in Fig.~\ref{fig:fig2}b show that SH radiation from the asymmetric-gap antenna dominates over its symmetric counterpart and exhibits a nearly ten-fold enhancement of radiated SH power, clearly demonstrating the substantial impact of local symmetry breaking on SHG. There remains a small amount of SH yield from the symmetric-gap antenna, which is not expected from the symmetric field enhancement in the gap and therefore vanishing DOA. As we show below, such residual SHG can be attributed to the presence of the so far neglected glass substrate, causing SH radiation due to symmetry breaking along the direction normal to the substrate.
	
\subsection{Quantitative modeling of second harmonic generation}

\begin{figure*}
	    \centering
	    \includegraphics[width=.99\linewidth]{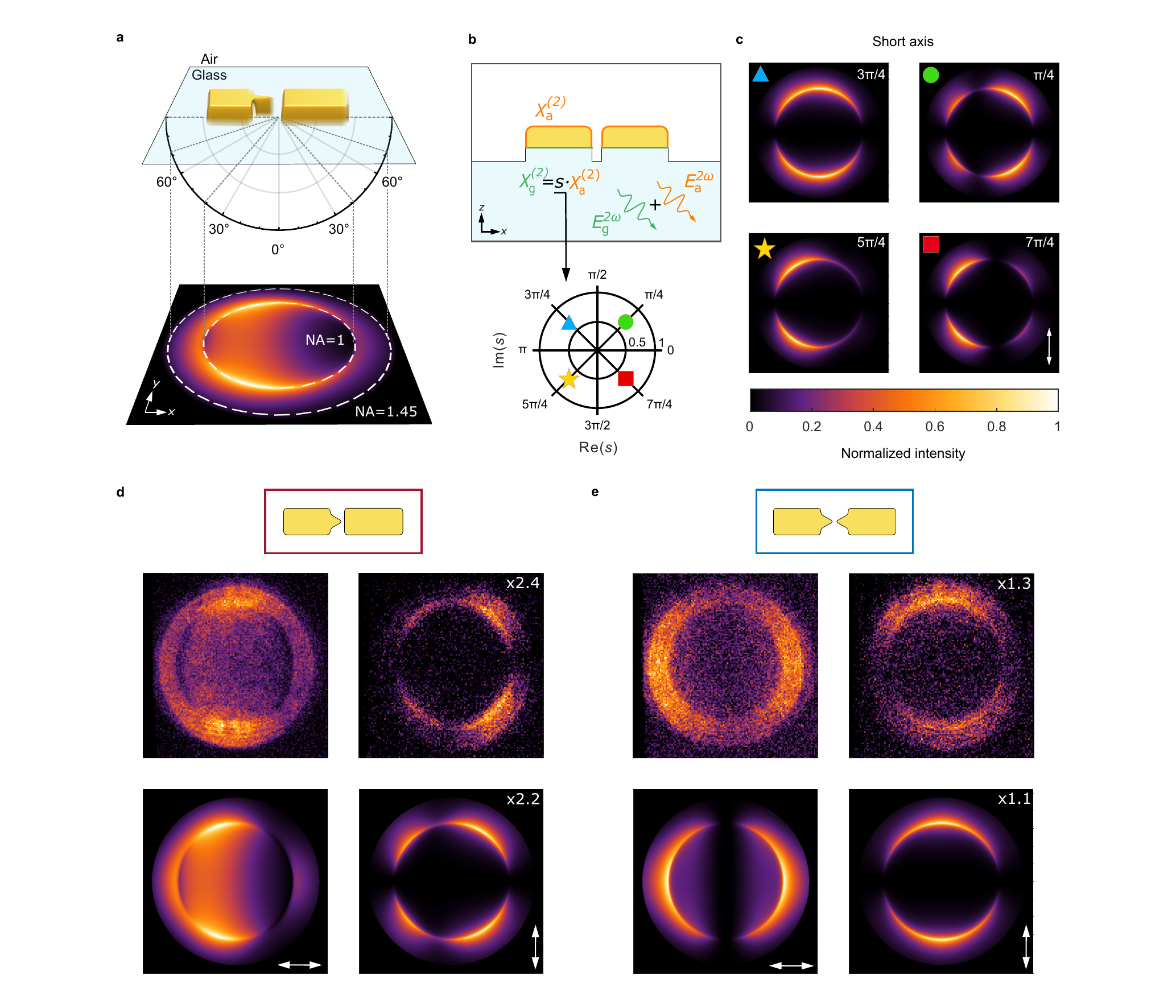}
	    \caption{\textbf{Simulated and experimental SH radiation patterns.} \textbf{a,}~Sketch of the emergence of a simulated radiation pattern by projecting the angular radiation pattern onto a plane in the substrate half-space. The critical angle of the glass-air interface and maximum collection angle for a numerical aperture of $1.45$ are indicated with white dashed lines. \textbf{b,}~Schematic $xz$-cut through the two antenna rods labeling the two distinct interfaces with corresponding second order susceptibility components, namely gold-air, $\chi_a^{(2)}$, and gold-glass, $\chi_g^{(2)}$. Lower panel in \textbf{b}: Complex plane representation of $s$. \textbf{c,}~Exemplary short axis radiation patterns for an asymmetric-gap antenna. The markers refer to the complex values of $s$ shown in the lower panel in \textbf{b}. \textbf{d,e,}~Experimental (top) and simulated (bottom) radiation patterns polarized along the long (left panel) and short (right panel) axis for an asymmetric-gap and symmetric-gap antenna, respectively, with $\alpha=40^\circ$. All radiation patterns are normalized to the maximum value of the corresponding long axis pattern. Simulations were performed with $|s|=0.55$ and $\phi_s=5/18\pi$\,rad.}
	    \label{fig:fig3}
	    \end{figure*}

\noindent In order to obtain a profound understanding of the SHG process and to validate the calculated values of the field enhancement and DOA presented in Fig.~\ref{fig:fig1}, we implemented nonlinear simulations based on a finite element method~\cite{celebrano_mode_2015}, that quantitatively describe the SH yield (for more details see Methods and Supplementary Section~1.3).
One major challenge in the modeling process is the specific choice of $\chi^{(2)}_{\text{S},\perp,\perp,\perp}$, which determines the strength of the SH polarization, see Eq.~\eqref{eq:P_s}. Various values can be found in literature, either based on theoretical calculations \cite{rudnick_second-harmonic_1971,sipe_analysis_1980,maystre_nonlinear_1986,benedetti_engineering_2010,antonietta_vincenti_gain-assisted_2012} or experimental data \cite{bachelier_origin_2010,van_nieuwstadt_strong_2006,krause_optical_2004,wang_surface_2009,boroviks_anisotropic_2021}. For instance, Wang \textit{et al.}~\cite{wang_surface_2009} used two-beam SHG on a sputtered gold film to retrieve the bulk and surface components of $\rttensortwo{\chi}^{(2)}_S$ for a gold-air interface. These values are commonly applied to model nanoscale systems~\cite{papoff_coherent_2015,obrien_predicting_2015,deng_giant_2020}, which typically results in a sufficient qualitative description of SHG. However, we find that none of the above-mentioned values lead to a satisfactory fit to our data (see Supplementary Section~1.4). Moreover, since all antennas are fabricated on a glass substrate, we have to take into account two distinct complex values of the SH susceptibility, $\chi^{(2)}_a$ for the gold-air and $\chi^{(2)}_g$ for the gold-glass interface. To overcome the ambiguity in the choice of the SH susceptibilities, we recorded SH radiation patterns (see Fig.~\ref{fig:fig3}a) carrying information about the dipole distribution and orientation as well as amplitude and phase of the SH sources, which in turn strongly depends on the relative strengths of $\chi^{(2)}_a$ and $\chi^{(2)}_g$ (for a detailed discussion see Supplementary Section~1.4).

We model our data by introducing a complex scaling parameter $s=|s|e^{i\phi_s}$, where $\chi^{(2)}_g=s\cdot \chi^{(2)}_a$, while $\chi^{(2)}_a$ is set to $1$ (see Fig.~\ref{fig:fig3}b). In the first step we calculate radiation patterns for a limited region in the complex plane with amplitude $|s|$ being scaled between 0 and 2 and phase $\phi_s$ between 0 and $2\pi$. In the following, we demonstrate the strong influence of $\phi_s$ on the symmetry of the radiation pattern, while we fix $|s|$ for all simulations shown in Fig.~\ref{fig:fig3} to 0.55 -- we show later that this exemplary value is an optimal choice. We focus on simulated radiation patterns polarized along the short axis of an asymmetric-gap antenna (\textit{i.e.}~along the $y$-direction in Fig.~\ref{fig:fig3}a). Depending on $\phi_s$, radiation patterns differ immensely (see Fig.~\ref{fig:fig3}c), which makes it possible to determine a coarse range of fitting values of $\phi_s$ by comparing these simulations to experimentally recorded radiation patterns.

\subsection{Experimental radiation patterns}

\noindent Experimental radiation patterns of an asymmetric-gap and symmetric-gap antenna are obtained by polarization-resolved back focal plane imaging. SH radiation patterns for both long and short axis polarization are displayed in the upper rows of Fig.~\ref{fig:fig3}d and e, respectively. The differences in the obtained radiation patterns suggest a distinct character of SH sources: For the symmetric-gap antenna, long and short axis signals are nearly equal and added up they form a rotationally symmetric radiation pattern. Such a radiation pattern is expected for a dipole that is oriented vertically with respect to the substrate, \textit{i.e.}~along the $z$-direction in Fig.~\ref{fig:fig3}b~\cite{bouhelier_near-field_2003}. Thus, for the symmetric-gap antenna, SHG mainly originates from the symmetry breaking due to the substrate. For the asymmetric-gap antenna, symmetry breaking along the long axis of the antenna is dominant. This expectation is indeed reflected in the radiation pattern that resembles that of an $x$-oriented SH dipole, being only slightly distorted by a small contribution from the symmetry breaking along $z$. In Supplementary Section~1.5 we discuss a semi-analytical approach to calculate these radiation patterns that further supports our model of the SHG process.

Comparison of the experimentally obtained short axis radiation pattern of the asymmetric-gap antenna to the corresponding simulated radiation patterns in Fig.~\ref{fig:fig3}c clearly shows that only for the first quadrant value of $\phi_s$ simulated and experimental radiation patterns match each other. Further analyzing the shape and intensity ratios between long and short axis radiation patterns in more detail results in a limited set of points for $s$ for which simulation and experiment are consistent (see Supplementary Fig.~S5). The lower row in Fig.~\ref{fig:fig3}d and e shows a complete series of simulated radiation patterns for $s=0.55e^{i5/18\pi}$. For this particular choice of $s$, which is optimal as fully explained later, we achieve excellent agreement between experimental and simulated radiation patterns for both asymmetric-gap and symmetric-gap antenna indicating that our model accurately describes the SH source.
		 
\subsection{Tuning of the SH efficiency}

\begin{figure}
	    \centering
	    \includegraphics[width=.99\linewidth]{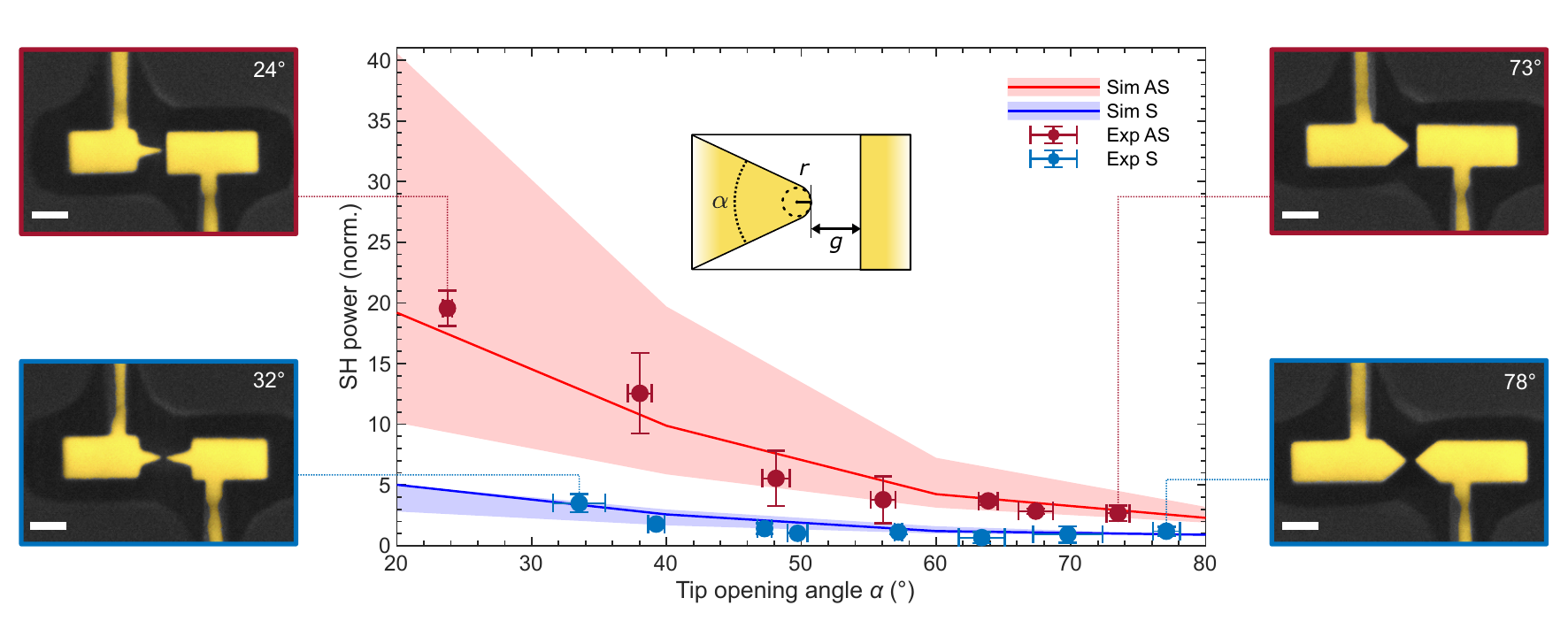}
	    \caption{\textbf{Tuning SHG by varying the degree of local symmetry breaking.} Measured and simulated SH power from asymmetric-gap (red) and symmetric-gap (blue) antennas as a function of the tip opening angle $\alpha$, \textit{i.e.} the degree of local asymmetry. Error bars in the experimental data account for measurements recorded for different nominally equal antennas with comparable linear optical properties. SEM images show antennas with smallest (left panel) and largest (right panel) $\alpha$ (scale bars, $50\,$nm). The inset defines additional gap parameters besides $\alpha$, namely gap size $g$ and radius of curvature of the tip $r$. Solid lines represent simulations with fixed gap size ($g=8.1\,$nm and $g=9.3\,$nm for asymmetric-gap and symmetric-gap antennas, respectively) and adjusted $r$ according to SEM analysis (see main text). The shaded regions illustrate simulations with $\Delta g,\Delta r=\pm 1\,$nm. Experimental and simulated data are separately normalized to the respective SH power of the symmetric-gap antenna with largest $\alpha$.}
	    \label{fig:fig4}
	    \end{figure}

\noindent To investigate the dependence of the SH efficiency on the DOA, effectively tuned by the tip opening angle $\alpha$ as shown in Fig.~\ref{fig:fig1}g and h, we performed a series of measurements of asymmetric-gap as well as corresponding symmetric-gap antennas with gradually decreasing $\alpha$ from $80^\circ$ down to $25^\circ$. Fig.~\ref{fig:fig4} displays the resulting radiated SH power as function of $\alpha$, as well as SEM images of antennas with smallest and largest angle. To guarantee comparability down to the involved crystal facets, all antennas were fabricated from the same gold microplatelet (thickness $40\,$nm) and only antennas with a resonance scattering amplitude $A_{\text{scatt}}(\lambda_\text{res})$ close to the design wavelength of $840\,$nm, specified by the condition $A_\text{scatt}(840\,\text{nm})/A_{\text{scatt}}(\lambda_\text{res})>0.9$, were taken into account (see Supplementary Section~2.3). For each antenna geometry the SH power of at least two nominally identical individual antennas was recorded. The SH power plots show that asymmetric-gap antennas always outperform their symmetric counterparts, especially for small $\alpha$, \textit{i.e.}~for the strongest local symmetry breaking. Notably, SHG from symmetric-gap antennas also slightly increases for smaller $\alpha$.

We apply our numerical model to simulate the tip opening angle dependence of the SH power. The electromagnetic field is strongly concentrated over the antenna tips which is why especially the gap size $g$ as well as the radius of curvature of the tip $r$ (see inset in Fig.~\ref{fig:fig4}) have a major influence. In experiments it is inevitable that $r$ decreases for smaller angles and it is therefore reasonable to adjust it also for the simulated geometries based on SEM analysis, while $g$ is kept constant at average values of $8.1\,$nm and $9.3\,$nm for asymmetric-gap and symmetric-gap antennas, respectively (see Supplementary Section~2.4 for a detailed analysis of the geometry parameters).
We have seen that the shape of simulated radiation patterns strongly depends on the phase $\phi_s$. The information about the tip angle dependence of the SH power contained in Fig.~\ref{fig:fig4} allows us to further confine the range of both, $\phi_s$ and $|s|$. To this end, we simulated the tip angle dependent SH power with $s$ ranging within the predetermined set of possible values based on the previous analysis of radiation patterns. Employing a fitting procedure, discussed in detail in Supplementary Section~1.4, we obtain an optimal value of $s=0.55e^{i5/18\pi}$. Both simulated radiation patterns (Fig.~\ref{fig:fig3}d and e) as well as tip angle dependent radiated SH power (solid lines in Fig.~\ref{fig:fig4}) fit well to our experimental data. Based on this remarkable quantitative agreement we can conclude that the simulated field enhancement values displayed in Fig.~\ref{fig:fig1}g and h accurately describe the experimental field distribution in the antenna gaps, and that the derived quantitative values for the DOA correctly reflect the asymmetry in the field hotspots.
	
To assess the influence of small-scale changes in the gap geometry on the SH efficiency, we further vary the geometry parameters $g$ and $r$ with $\Delta g,\Delta r = \pm1\,$nm, an uncertainty that follows from the inspection of SEM images. Maximum and minimum values of the shaded regions in Fig.~\ref{fig:fig4} correspond to simulations, where $g$ and $r$ are equally modified, \textit{i.e.}~$\Delta g,\Delta r = -1\,$nm for maximum and $\Delta g,\Delta r = +1\,$nm for minimum SH power, respectively. For symmetric-gap antennas, varying these parameters does not impact the SH power too much, as the symmetry in the gap is not lifted. Asymmetric-gap antennas, on the contrary, are extremely sensitive to changes in the gap region with more than a $100\%$ boost in SH power for the smallest tip opening angle of $20^\circ$. This is also reflected in experiments, where we observe a larger variation of SH power for nominally equal asymmetric-gap antennas, especially towards smaller $\alpha$. Nevertheless, nearly all of the investigated antennas lie within the range of $\Delta g,\Delta r = \pm1\,$nm, which demonstrates the outstanding control and precision we are able to achieve with the presented combined Ga-/He-FIB approach.

\subsection{Conclusion}

\noindent  We have realized efficient SHG from plasmonic dimer antenna hotspots by controlling both enhancement and asymmetry of local fields in sub-10$\,$nm gaps. Such local symmetry breaking preserves the large field enhancement of the resonant mode at the FF and circumvents the silencing effect. The maximal observed nonlinear coefficient ($\gamma_\text{SH}=1.7\times10^{-10}\,\text{W}^{-1}$, see Supplementary Section~2.5) is comparable to the coefficient observed in systems with two resonances ($\gamma_\text{SH}=5.1\times10^{-10}\,\text{W}^{-1}$), where symmetry is broken globally \cite{celebrano_mode_2015}. By employing He-FIB milling of mono-crystalline gold microplatelets we achieve such high precision and reproducibility that we are able to systematically vary the antenna gap geometry. It is therefore possible to gradually tune the SH radiation by adjusting the opening angle of the sharp tip at the antenna gap. 

Recorded SH radiation patterns reveal the orientation and phase relation of SH dipoles determined by the relative strengths of the second-order surface susceptibilities of gold-air and gold-glass interfaces. Full-wave nonlinear simulations result in a simultaneous quantitative matching between simulated and experimentally observed radiation patterns as well as tip angle dependent SH power for both asymmetric-gap and symmetric-gap antennas.

Based on the quantitative analysis we conclude that we reach a field enhancement of up to 175 and a DOA of up to 0.36 in our experiments. Although in similar systems comparable asymmetries can be obtained, field enhancements are an order of magnitude lower~\cite{karnetzky_towards_2018,zimmermann_toward_2019,shi_femtosecond_2021}. We therefore anticipate that our concept of local symmetry breaking opens new possibilities for the realization of efficient nonlinear nanoscale systems, which demand strong field enhancement combined with large local field asymmetry.

\subsection{Methods}
\subsubsection{Sample fabrication}
Mono-crystalline gold microplatelets with a thickness of $40\,$nm are grown based on a wet-chemical synthesis described in Ref.~\cite{krauss_controlled_2018} and are afterwards transferred onto a glass slide with evaporated metal layers featuring an array of circular openings produced by electron beam lithography ($20\,$nm chromium adhesion layer, $80\,$nm gold layer), see Ref.~\cite{kern_electrically_2015} . The microplatelets are placed on top of sufficiently small holes to ensure conductive connection of flake and metal film. Plasmonic nanoantennas are fabricated using a three-step Ga-/He-FIB (\textit{ZEISS Orion Nanofab}) approach described in Supplementary Section~2.1. 
	
\subsubsection{Optical characterization}
To record dark-field white light scattering spectra of asymmetric-gap and symmetric-gap plasmonic nanoantennas, the sample is excited with a collimated beam from a stabilized halogen lamp (Thorlabs SLS201L/M) using an oil-immersion microscope objective (PlanApochromat, 100×, NA = 1.45, Nikon). The scattered light is collected via the objective and a circular beam block is used to block direct reflection from the sample. The scattered light is then analyzed by a spectrometer (Shamrock 303i, 80 lines/mm, blazing at 870 nm) equipped with an electron-multiplied charge-coupled device (EMCCD, iXon A-DU897-DC-BVF, Andor).

For SHG measurements a femtosecond pulsed Ti:Sa laser (Mira Optima 900-F, repetition rate $80\,$MHz, pulse width at laser ouput $100\,$fs) is focused through the glass substrate on single antennas, using the same objective and spectrometer as above. Details of the setup and the implemented pulse compression system can be found in the Supplementary Section~2.2. The SH spectra displayed in Fig.~\ref{fig:fig2} have been recorded using a 300 lines/mm grating blazed at $420\,$nm with an integration time of $60\,$s, while for recording of the SH radiation in Fig.~\ref{fig:fig4} we used an 80 lines/mm grating blazed at $565\,$nm and an integration time of $30\,$s. All spectra are corrected with the total transfer function of the detection path (see Supplementary Section~2.5). To record SH radiation patterns we insert a 1000-mm Betrand lens before the spectrometer and replace the spectrometer grating by a mirror. Integration times were set to $120\,$s for asymmetric-gap and $150\,$s for symmetric-gap antennas, respectively. 
	
\subsubsection{Numerical simulations}
We perform frequency-domain finite-element simulations to calculate the linear and nonlinear response of the system by using the commercial software package COMSOL Multiphysics. The source employed at the SH (Eq.~\eqref{eq:P_s}) is calculated from linear simulations where the structure is excited from the substrate with a plane wave polarized along the long axis of the antenna. The mesh is generated to match the symmetry of the simulated structure, with a minimum element size of $0.5\,$nm. SH radiation patterns and SH power values are obtained from far-field calculations using the Matlab toolbox RETOP 8.1~\cite{yang2016near}, with the near-field collected at the surface of a box $100\,$nm from the structure. More details about the numerical simulations can be found in the Supplementary Section~1.

\begin{acknowledgement}

We gratefully acknowledge funding by the Deutsche Forschungsgemeinschaft (DFG, German Research Foundation) under Germany’s Excellence Strategy through the Würzburg-Dresden Cluster of Excellence on Complexity and Topology in Quantum Matter ct.qmat (EXC 2147, Project ID ST0462019) as well as through a DFG project (INST 93/959-1 FUGG), a regular project (HE5618/10-1), and a Reinhard Koselleck project (HE5618/6-1). The Volkswagen foundation is acknowledged for funding via the Experiment! program (95869). Thorsten Feichtner additionally acknowledges financial support from the European Commission through the Marie Skłodowska-Curie Actions (MSCA) individual fellowship project PoSHGOAT (project-
id 837928) and participation in CA19140 (FIT4NANO), supported by COST (European Cooperation in Science and Technology). 

The authors thank Marco Finazzi, Michele Celebrano and Paolo Biagioni (all Politecnico Milano) as well as Constantino De Angelis (University of Brescia) for helpful discussions. Furthermore, the authors thank Monika Emmerling and Patrick Pertsch (both University of Würzburg) for substrate preparation and microplatelet transfer.

\end{acknowledgement}

\begin{suppinfo}

Details about numerical simulations: geometry setup, linear simulations, overview over different approaches for nonlinear simulations, analysis of the influence of $\chi^{(2)}$ on radiation patterns, effective dipole model;
Experimental information: three-step milling method, scheme of optical setup, summary linear measurements, analysis of antennna geometries with SEM and AFM, evaluation of efficiency

\end{suppinfo}

\bibliography{SHG}

\end{document}